\documentclass[sn-mathphys,Numbered, pdflatex]{sn-jnl}% Math and Physical Sciences Reference Style
\usepackage{graphicx}%
\usepackage{multirow}%
\usepackage{amsmath,amssymb,amsfonts}%
\usepackage{amsthm}%
\usepackage{mathrsfs}%
\usepackage[title]{appendix}%
\usepackage{xcolor}%
\usepackage{textcomp}%
\usepackage{manyfoot}%
\usepackage{booktabs}%
\usepackage{algorithm}%
\usepackage{algorithmicx}%
\usepackage{algpseudocode}%
\usepackage{listings}%

\usepackage{siunitx}
\DeclareSIUnit\pixel{px}
\usepackage{float}
\usepackage{physics}
\usepackage{chemformula}
\usepackage{upgreek}
\usepackage{svg}
\usepackage{layouts}

\begin{document}

\title[Interactions between flow fields induced by surface dielectric barrier discharge arrays]{Interactions between flow fields induced by surface dielectric barrier discharge arrays}

%%=============================================================%%
%% Prefix	-> \pfx{Dr}
%% GivenName	-> \fnm{Joergen W.}
%% Particle	-> \spfx{van der} -> surname prefix
%% FamilyName	-> \sur{Ploeg}
%% Suffix	-> \sfx{IV}
%% NatureName	-> \tanm{Poet Laureate} -> Title after name
%% Degrees	-> \dgr{MSc, PhD}
%% \author*[1,2]{\pfx{Dr} \fnm{Joergen W.} \spfx{van der} \sur{Ploeg} \sfx{IV} \tanm{Poet Laureate} 
%%                 \dgr{MSc, PhD}}\email{iauthor@gmail.com}
%%=============================================================%%

\author*[1]{\fnm{Alexander} \sur{Böddecker}}\email{boeddecker@aept.rub.de}

\author[2]{\fnm{Maximilian} \sur{Passmann} }
% \equalcont{These authors contributed equally to this work.}

\author[1]{\fnm{Sebastian} \sur{Wilczek} }

\author[1,3]{\fnm{Lars} \sur{Schücke} }

\author[1]{\fnm{Ihor} \sur{Korolov} }

\author[2]{\fnm{Romuald} \sur{Skoda} }

\author[1]{\fnm{Thomas} \sur{Mussenbrock} }

\author[3]{\fnm{Andrew R.} \sur{Gibson} }

\author[1]{\fnm{Peter} \sur{Awakowicz} } %\email{awakowicz@aept.rub.de}

% \equalcont{These authors contributed equally to this work.}

\affil*[1]{\orgdiv{Chair of Applied Electrodynamics and Plasma Technology}, \orgname{Ruhr University Bochum}, \orgaddress{\street{Universitätsstraße 150}, \city{Bochum}, \postcode{44801}, \country{Germany}}}

\affil[2]{\orgdiv{Chair of Hydraulic Fluid Machinery}, \orgname{Ruhr University Bochum}, \orgaddress{\street{Universitätsstraße 150}, \city{Bochum}, \postcode{44801}, \country{Germany}}}

\affil[3]{\orgdiv{Research Group for Biomedical Plasma Technology}, \orgname{Ruhr University Bochum}, \orgaddress{\street{Universitätsstraße 150}, \city{Bochum}, \postcode{44801}, \country{Germany}}}
\abstract{This study investigates the flow field induced by a surface dielectric barrier discharge (SDBD) system, known for its efficient pollution remediation of volatile organic compounds (VOCs). We aim to understand the flow dynamics that contribute to the high conversion observed in similar systems. Experimental techniques, including schlieren imaging and particle image velocimetry (PIV), applied with high temporal resolution, were used to analyse the flow field. Complementary, fluid simulations are employed to investigate the coupling between streamer and gas dynamics. 
Results show distinct fluid field behaviours for different electrode configurations, which differ in geometric complexity. The fluid field analysis of the most basic electrode design revealed behaviours commonly observed in actuator studies. The simulation results indicate the local information about the electron density as well as different temporal phases of the fluid flow. The electrode design with mostly parallel grid line structures exhibits confined vortices near the surface. In contrast, an electrode design also used in previous studies, is shown to promote strong gas transport through extended vortex structures, enhancing gas mixing and potentially explaining the high conversion observed.}


%%\pacs[JEL Classification]{D8, H51}

%%\pacs[MSC Classification]{35A01, 65L10, 65L12, 65L20, 65L70}

\maketitle

\section{Introduction}
\label{chap:introduction}

The control of volatile organic compound (VOC) emissions has become a crucial task for preserving of the environment and safeguarding human health. Exposure to VOCs has been linked to various diseases, including cancer, birth defects in children, neurodevelopmental disorders, cardiovascular diseases and asthma 
\cite{pereraEarlylifeExposurePolycyclic2014, cosselmanEnvironmentalFactorsCardiovascular2015, sarigiannisExposureMajorVolatile2011, heusinkveldNeurodegenerativeNeurologicalDisorders2016}.

In contrast to conventional methods such as thermal oxidizers, non-thermal atmospheric pressure plasma (NTP) systems, particularly dielectric barrier discharges (DBDs), have emerged as promising technologies for VOC pollution control. NTP systems offer several advantages, including independence from fossil fuels and fast response times. Among NTPs, DBDs exhibit high energy efficiencies and scalability, making them a suitable discharge source for VOC pollution control systems. The formed discharges inside a DBD contain highly energetic electrons (several eV) which generate reactive species that can oxidize VOCs.

Within the DBD category, two basic configurations exist: Volume DBDs (VDBDs) and surface DBDs (SDBDs). Both consist of two opposing electrodes and a dielectric barrier to prevent the transition to an arc discharge. VDBDs feature a gas-filled space between the electrodes, allowing ignition to occur within this region. On the other hand, SDBDs have the entire gap, where the ignition condition would be normally fulfilled, filled with a dielectric material, restricting ignition to the surface of the electrodes. SDBDs have found more frequent application in chemical and aerodynamic manipulation of flowing media due to their potential to influence larger air volumes through discharge positioning.

The term “plasma actuator” has been used in plasma aerodynamics for over two decades to describe a specific type of SDBD used for active flow control  \cite{corkeDielectricBarrierDischarge2010,poggie2015}. These actuators employ highly asymmetric electrode geometries in flow direction, differing from the common parallel plate geometries, which result in an asymmetric plasma discharge occurring only on one side of the electrode. When the discharge is ignited, the ionized gas acts as a body force to the ambient neutral air, commonly referred to as ionic wind \cite{robinsonHistoryElectricWind1962a}. This body force is utilized to actively manipulate aerodynamics \cite{greenblatt2012,feng2015}, such as controlling boundary layer separation over airfoils at low Reynolds-numbers \cite{sato2015}.
Boeuf et al. \cite{boeuf2007electrohydrodynamic} studied the ionic wind in the context of electrohydrodynamic force by 2D fluid simulations in SDBDs. Zhang et al. \cite{zhang2021computational} used kinetic 2D particle-in-cell / Monte Carlos collisions (PIC/MCC) simulations to investigate the propagation mechanisms of plasma streamers with a nanosecond rise time square voltage pulse. 

In this study, we investigate three different twin SDBD electrode configurations with varying geometric complexities to facilitate fundamental research and validate the used diagnostic methods. One of the electrode configurations has been predominantly studied in previous works regarding its VOC conversion performance and fundamental discharge parameters. Offerhaus \emph{et al.} \cite{offerhausSpatiallyResolvedMeasurements2017b} and Kogelheide \emph{et al.} \cite{kogelheideCharacterisationVolumeSurface2020} measured electron density distributions and reduced electric field strengths using optical emission spectroscopy (OES). The influence of $\mu$s or ns voltage excitation towards electrode erosion, dissipated power, and effective discharge parameters is investigated by Ngyuen-Smith \emph{et al.} \cite{nguyen-smithMsNsTwin2022}. Schücke \emph{et al.} \cite{schuckeConversionVolatileOrganic2020} conducted research on the VOC conversion of multiple VOCs in combination with gas chromatography-mass spectrometry and power dissipation with the SDBD, as well as the study of nitrogen and oxygen species densities generated by the SDBD and their influence on conversion using optical absorption spectroscopy \cite{schuckeOpticalAbsorptionSpectroscopy2022}. These reactive species density studies were further complemented by a zero-dimensional chemistry model by Schücke \emph{et al.} \cite{schuckeAnalysisReactionKinetics} The influence of an additional $\upalpha$-\ch{MnO2} catalyst coated on the surface of the SDBD is shown by Peters \emph{et al.} \cite{petersCatalystenhancedPlasmaOxidation2021} to the carbon balance, $n$-butane conversion, and selectivity. Böddecker \emph{et al.} \cite{boddeckerScalableTwinSurface2022b} have shown the potential of a multiple electrode SDBD system for VOC conversion at high gas flow rates, with additional studies focusing on the removal of ozone as a potentially harmful by-product of the discharge.

The conversion efficiency measured by Böddecker \emph{et al.} \cite{boddeckerScalableTwinSurface2022b} with a scaled-up SDBD system was found to be comparatively high, and they suggested that complex fluid mechanics may be the underlying reason. Vortices, which are also observed in plasma actuator studies, could interact, resulting in a velocity field that enhances the transport of VOC molecules into the active plasma region close to the surface.

The fluid mechanical analysis of the SDBD is a promising way to further optimize its performance towards the VOC conversion and its underlying gas chemistry. The literature on plasma actuators has revealed different behaviours, and a study by Whalley and Choi \cite{whalleyStartingTravelingColliding2010} visually demonstrated the generation and qualitative development of vortices induced by a plasma actuator. During the initial phase after DBD ignition, a starting vortex is generated, which grows in size and moves along the surface. The interaction between vortices generated by two opposing plasma actuators leads to motion normal to the surface \cite{whalleyStartingTravelingColliding2010} and the formation of a laminar wall jet \cite{jukesCharacterizationSurfacePlasmaInduced2006}.  In a subsequent investigation, Whalley and Choi \cite{whalleyStartingVortexQuiescent2012} demonstrated that their plasma actuator exerted a constant force on the fluid, resulting in a steady state fluid velocity. Dickenson \emph{et al.} studied the spatio-temporal distribution of reactive nitrogen species generated from a SDBD using particle image velocimetry, laser induced fluorescence and numerical modelling. They showed how the aerodynamics influence the reactive species transport, which could be substantial for gas treatment processes \cite{dickensonGenerationTransportReactive2018a}. Gilbart \emph{et al.} did research on the influence of the electrode width of a SDBD to the dissipated power, the induced flow and also to reactive species distributions, by applying their numerical model as well as particle image velocimetry. They observed a transition between a convection-driven and diffusion-driven state which influences the reactive species distribution strongly \cite{gilbartMutualInteractionMultiple2022}.

In a recent study, Ollegott \emph{et al.} \cite{ollegott2023} investigated the flow characteristics in an SDBD system using schlieren imaging. They focused on the conversion of \ch{O2} traces in a gas mixture containing \ch{H2}. The researchers successfully observed schlieren structures throughout the entire volume of their reactor, providing evidence of mass flow induced by the discharge. Additional fluid dynamic simulations support their observations.

In our study, we compare time-resolved schlieren and particle image velocimetry (PIV) measurements for all three SDBD electrode configurations to reach a better understanding of the gas transport and exchange. The corresponding simulation for the basic electrode configuration computes the streamer dynamics as well as its impact to the fluid dynamics. This allows a deeper analysis of the responsible and dominant physical effects that mainly induce the fluid flow in this case. 

In contrast to the study of Ollegott \emph{et al.} \cite{ollegott2023} our measurements will be time-resolved, allowing for the extraction of velocity information also from the schlieren measurements. Supplementary simulation results show the induced flow of the SDBD, based on the underlying streamer dynamics. 

The remainder of this paper is structured as follows. Section \ref{chap:experiment} contains details of the SDBD setup and electrode configurations, measurement techniques, and simulation setup. This is followed in Sec. \ref{chap:results} by the presentation and discussion of the results, including an electric characterization of the SDBD, and a description of the transient start phase and steady-state flow field for the three different electrode configurations. After the experimental part, complementary simulation results for the basic electrode configuration are presented and discussed in Sec. \ref{chap:Simulation_Results}. Finally, the main conclusions and an outlook for future work are presented in Sec. \ref{chap:conclusion}.

\section{Experiment and diagnostics}
\label{chap:experiment}

\subsection{Surface dielectric barrier discharge}

\begin{figure*}[hbt]
    \centering
    \includegraphics[width=0.95\textwidth]{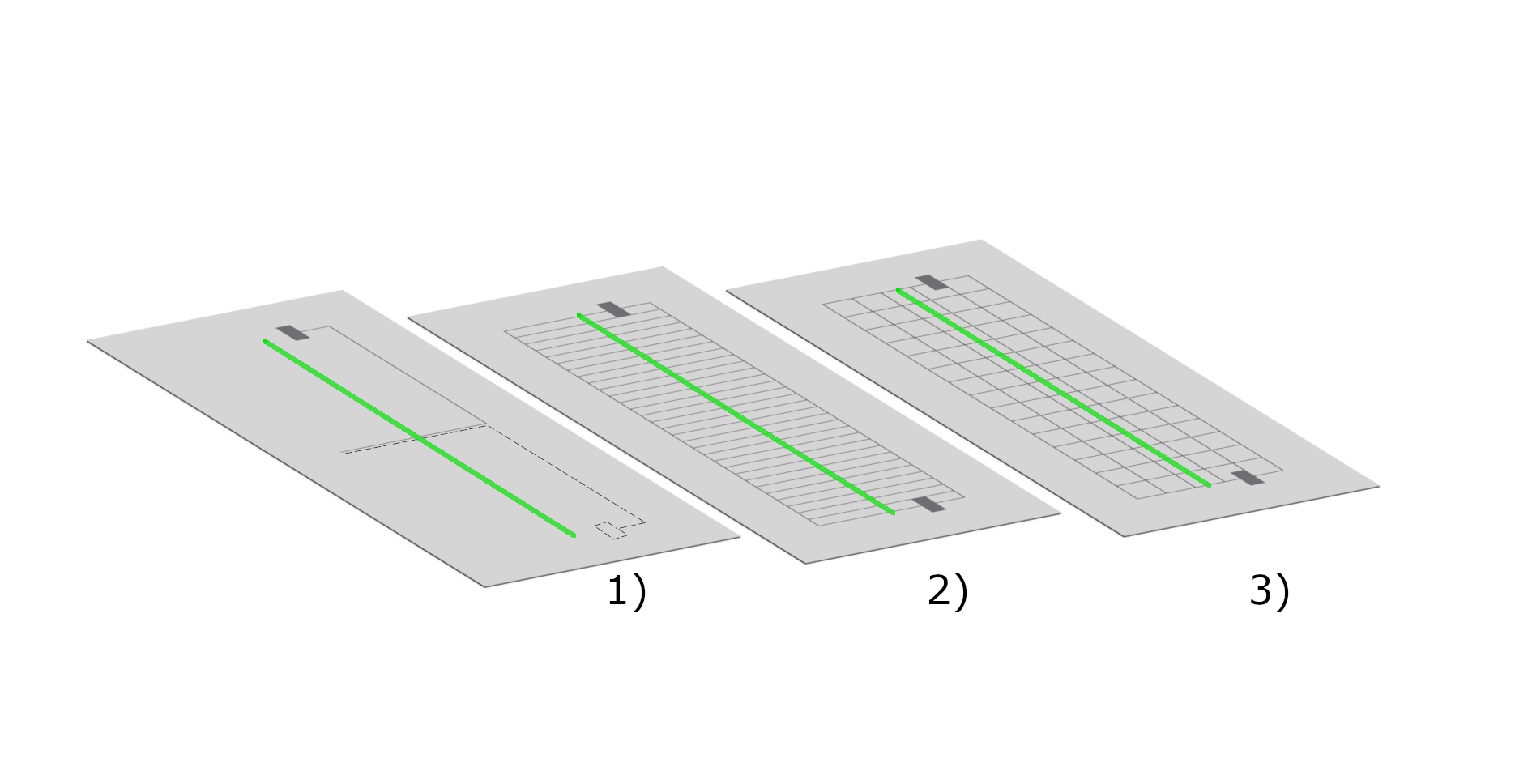}
	\caption{Schematic drawing of the three SDBD electrode configurations used in this work. The single grid line electrode 1), the parallel grid line electrode 2) and the squared grid line electrode 3) are shown. Exclusively for configuration 1) the counter electrode on the bottom side of the plate is printed asymmetrically (dashed line illustrates grid on the bottom side), that the grid lines only overlap at the centre of this configuration. The green lines along the centre of the electrode configurations represent the positions of the laser sheet for the PIV diagnostic, which is described in more detail in section \ref{PIV fundamentals}.}
	\label{fig:electrodes}
\end{figure*}

Three different SDBD electrode configurations (fabricated by Alumina Systems GmbH) are used for the investigations. Each configuration consists of a dielectric plate, which is made of $\upalpha$-\ch{Al2O3} with a size of \SI{190}{mm} x \SI{88}{mm} x \SI{0.635}{mm}. Two conducting metal grids, each one screen printed on both sides of the ceramic plate, serve as the interacting counter electrodes. The metal grid is manufactured out of molybdenum (\SI{80}{\%}) as well as manganese silicate (\SI{20}{\%}) and is nickel plated chemically. A minimum distance between the grid and the edge of the plate of \SI{18.8}{mm} prevents the discharge from igniting along the surface towards its corresponding counter electrode directly.

Figure \ref{fig:electrodes} shows the three SDBD configurations used in this study. While configurations 1) and 2) are specially designed for the investigation of the induced fluid flow in this study, configuration 3), which is also called "squared grid line electrode", was widely used in previous investigations, mainly motivated for use in gas-phase plasma-chemical conversion applications \cite{offerhausSpatiallyResolvedMeasurements2017b, kogelheideCharacterisationVolumeSurface2020, nguyen-smithMsNsTwin2022, schuckeConversionVolatileOrganic2020, schuckeOpticalAbsorptionSpectroscopy2022, schuckeAnalysisReactionKinetics, petersCatalystenhancedPlasmaOxidation2021, boddeckerScalableTwinSurface2022b}.

% This configuration has a total grid area of \SI{150}{mm} x \SI{50}{mm} per side and a quadratic lattice constant of \SI{10}{mm}. The metallic grid lines are \SI{0.45}{mm} wide. For configuration 2), which is called "parallel grid lines electrode", it is expected, to show a more basic fluid mechanical behaviour because most of the perpendicular crossing grid lines are missing and the 2-dimensional analysis will lead to a better resolution of the actual processes taking place with the applied diagnostics. The distance between the grid lines is \SI{5}{mm}, which allows to investigate the impact of other distances between neighbouring grid lines to the induced flow by the SDBD at the same time. To reduce the complexity further, electrode configuration c), which is referred as "single grid line electrode", is investigated to study the uninfluenced induced flow without interacting fluid flows induced from neighboured grid lines. There, the grid is printed asymmetrically on both sides, so that only the grid lines in the middle of the plate have a counter electrode which is close enough to reach a sufficient electric field strength for the discharge ignition. Therefore, only two-dimensional effects are expected to be observed with electrode configuration c). \\

To reduce the geometric complexity and hence the complexity of the resulting fluid flow field from type 3), electrode configuration 1), which is referred to as a "single grid line electrode", is investigated to study the uninfluenced induced flow without interacting fluid flows induced from neighboured grid lines. The grid configuration is designed such that the grid lines are printed in a manner where only the grid lines in the middle of the plate overlap, ensuring the presence of a counter electrode in close proximity. Therefore, exclusively at the grid line in the middle, the ignition can take place. This is indicated by the dashed line in figure \ref{fig:electrodes}, which shows the position of the grid line on the bottom side of the plate. The metallic grid lines are \SI{0.45}{mm} wide. This electrode design is well-suited for two-dimensional simulations, assuming that for this case the flow field is not three-dimensional. Hence, it provides an ideal basis to validate the simulation results against the measurements and gain additionally fundamental insights. 
For configuration 2), which is referred to as the "parallel grid line electrode", parallel neighbouring grid lines are added, representing a test case of intermediate complexity. To maintain a similar capacitance relative to type 3), the distance between the grid lines is set to \SI{5}{mm}, which is half the value of that configuration.  
Configuration 3) is expected to develop the most complex and highly three-dimensional flow field, because of the perpendicular crossing grid lines and will be analysed with respect to the results of type 1) and 2). 

\subsection{Electrical supply}
As used in previous studies \cite{kogelheideCharacterisationVolumeSurface2020, offerhausSpatiallyResolvedMeasurements2017b, schuckeConversionVolatileOrganic2020,schuckeOpticalAbsorptionSpectroscopy2022,petersCatalystenhancedPlasmaOxidation2021, nguyen-smithMsNsTwin2022, schuckeAnalysisReactionKinetics, boddeckerScalableTwinSurface2022b} a high voltage generator (Plasma Generator G2000, Redline Technologies, Germany) was used. In this system, unipolar voltage pulses (maximum: \SI{300}{V}, \SI{500}{kHz}) in combination with a transformer and the natural capacitance of the SDBD are transformed into damped sinusoidal waveforms, by utilizing the resulting resonant circuit. The resonance frequency depends on the capacitance of the system and is about \SI{86}{kHz} for the squared grid electrode geometry 3). Because of the damping, the peak-to-peak voltage of the first two half waves is used to describe the waveform amplitudes. Voltages between \SI{8}{kV_{pp}}-\SI{13}{kV_{pp}} were used for gas conversion measurements in previous studies. Due to constructive limitations of the electric power supply the maximum possible repetition frequency is restricted to \SI{4}{kHz} \cite{schuckeConversionVolatileOrganic2020}. In this study, a voltage of \SI{11}{kV_{pp}} is used to allow a frequency variation between \SI{1}{kHz} to \SI{4}{kHz} for all electrode configurations. \\
While the voltage is measured with a high voltage probe (6015A, Tektronix GmbH, Germany), the current is measured with a current probe (Model 6585, Pearson Electronics, USA). Both signals are recorded by an oscilloscope (Waverunner 824, Teledyne LeCroy, USA) and the power $P$ can be calculated as following:

\begin{equation}
    P = f_{\mathrm{rep}}\int_{0}^{T_{\mathrm{p}}} U(t)\left( I(t) - C \: \dv{U}{t} \right) \; \mathrm{d}t
    \label{Eq:Power}
\end{equation}

Here $T_{\mathrm{P}}$ is the pulse duration, $f_{\mathrm{rep}}$ the repetition frequency, $U(t)$ the voltage waveform, $I(t)$ the current waveform and $C$ the capacitance of the system which goes into account for the calculation of the displacement current. The displacement current can be neglected to decrease the computation time for a real time process control, because its contribution is negligibly low. A more detailed description can be found in Schücke \emph{et al.} \cite{schuckeConversionVolatileOrganic2020} and Böddecker \emph{et al.} \cite{boddeckerScalableTwinSurface2022b}.

\subsection{Schlieren setup}

Density variations caused by the SDBD were visualized by means of schlieren photography. Schlieren methods are based on the relation between refractive index $n$ and the local density $\rho$, which in case of air is given by
\begin{align}
n-1 =  k \cdot \rho,
\end{align}
where $k$ is the Gladstone-Dale constant. Light rays are refracted by optical inhomogeneities, such as density variations, proportional to their gradients of refractive index \cite{settlesSchlierenShadowgraphTechniques2001}:
\begin{align}
\label{eq:ray_deflection}
\varepsilon_x = \int{\frac{1}{n}\frac{\partial n}{\partial x} \mathrm{d}z}, \, \varepsilon_y = \int{\frac{1}{n}\frac{\partial n}{\partial y} \mathrm{d}z}.
\end{align}
Herein, $x$ and $y$ denote the in-plane directions and $z$ is the direction normal to the image plane, which is aligned with the optical axis of the schlieren system. The resulting ray deflection angles are converted and amplified by the schlieren system into changes in brightness, i.e. image intensity.

\begin{figure}[b]
    \centering
    \includegraphics[width=0.55\linewidth]{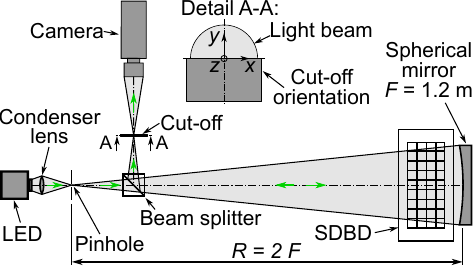}
	\caption{Schematic of the single-mirror coincident schlieren system.}
	\label{fig:schlieren_system}
\end{figure}
In the present study, a single-mirror coincident schlieren system was used \cite{taylorImprovementsSchlierenMethod1933}, which is shown schematically in Figure~\ref{fig:schlieren_system}. Light is emitted from a light emitting diode (LED, model: Luminus Devices SFT10, $\lambda=\SI{555}{\nano\meter}$) and focused with a collimating lens onto a pinhole to generate a point-like source of light. This light illuminates a spherical mirror which is placed at the radius of curvature, i.e. a distance of $R = 2F$, where $F$ refers to the focal length of the mirror. Because the sensitivity of a schlieren system is directly proportional to the focal length $F$ of the mirror \cite{settlesSchlierenShadowgraphTechniques2001}, the sensitivity of a single-mirror configuration is effectively twice that of more conventional setups, such as $z$-Type schlieren systems (see also Sec. \ref{sec:exp_uncertainty}). The reflected light passes the measurement section, where the SDBD is placed, twice and forms a source image on the light source. To separate this image from the light source, a beam splitter is inserted between the light source and the mirror to guide one part of the diverging beam to the camera (Phantom Veo-410). A knife edge in horizontal orientation (cf. figure~\ref{fig:schlieren_system}) is used in front of the camera to cut off the image at the focal point so that only light rays deflected by schlieren can pass the knife edge. The resulting image shows the phase differences of the light that has travelled through the fluid influenced by the SDBD. Due to the horizontal orientation of the knife-edge, only ray deflections in the $y$-direction, i.e. $\varepsilon_y$ (cf. Eq.~\ref{eq:ray_deflection}), are visible in the schlieren images. To obtain a field-of-view comparable to the PIV measurements, the schlieren system was set up such, that only the flow above the electrode is visualized. This field-of-view represents a compromise between an image size that depicts multiple electrode grid lines in the squared grid electrode case and a high spatial resolution to resolve the schlieren structures properly.

\subsection{Particle image velocimetry setup}
\label{PIV fundamentals}

\begin{figure}[b]
    \centering
    \includegraphics[width=0.5\linewidth]{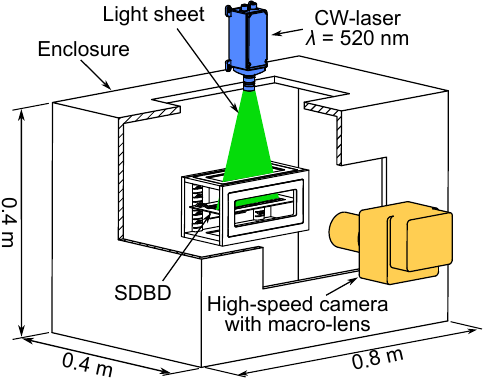}
	\caption{Schematic of the planar (2D-2C) particle-image velocimetry (PIV) setup.}
	\label{fig:PIV_system}
\end{figure}

Time-resolved planar (2D-2C) particle image velocimetry (PIV) was used to obtain detailed quantitative information about the velocity field in the vicinity of the SDBD. Figure \ref{fig:PIV_system} shows the schematic setup of the PIV system and its main components. The discharge chamber with a single SDBD installed was placed on a laboratory lifting table (not shown in figure~\ref{fig:PIV_system}) at the centre of a large enclosure having length width and height of $\SI{0.8}{\meter}$, $\SI{0.4}{\meter}$, and $\SI{0.4}{\meter}$ respectively, designed to contain the seeding particles and to shield the SDBD from large-scale random air movement inside the laboratory. Two large glass windows in the front and top sidewalls of the enclosure allowed for optical access.

Illumination of the flow-field was provided by a continuous wave (CW) diode laser (Z-LASER, model: ZQ1) with a wavelength of $\lambda=\SI{520}{\nano\meter}$ and a maximum output power of $\SI{800}{\milli\watt}$. The integrated light-sheet optics with adjustable focus produced a uniform light sheet with an in-plane opening angle of $\SI{30}{\degree}$ and a thickness of order $\SI{100}{\micro\meter}$. The flow was seeded with DEHS (Di-ethyl-hexyl-sebacate) droplets of approximately $\SI{1}{\micro\meter}$ in diameter using a commercially available aerosol generator (LaVision). After filling the enclosure with seeding particles, a waiting time of $\SI{10}{\min}$ was imposed before taking a measurement, to ensure that flow induced by the filling process had settled down. During that period, the laser was also switched off to avoid any thermally induced flow close to the SDBD surface caused by the intense laser light.

Particle images were recorded with a high-speed camera (Phantom VEO-410) at full sensor resolution of $1280\times\SI{800}{\square\pixel}$ using a $\SI{100}{\milli\meter}$ $f2.8$ manual focus macro-lens. Images were recorded at a frame-rate of $\SI{5200}{fps}$ for all configurations. However, for the two lower repetition rates of $f=\SI{1}{\kilo\hertz}$ and $\SI{2}{\kilo\hertz}$ the temporal resolution was later sampled down by a factor of two during post-processing to obtain a sufficient particle displacement of less than $1/4$ of the final interrogation window size, i.e. $<\SI{6}{pixel}$ \cite{scharnowski2020,raffel2018}.

Post-processing of the image data was performed with the well established open-source toolbox PIVlab \cite{thielicke2014,thielicke2021} (version 2.61) for Matlab. First, the raw images were  pre-processed using contrast limited adaptive histogram equalization (CLAHE), which is known to significantly improve cross-correlation results \cite{shavit2007}. The individual image pairs were then analysed using a multi-pass cross-correlation approach with FFT window deformation. In total, three interrogation window sizes of $64\times64$, $32\times32$, and $24\times \SI{24}{pixel^2}$ with $\SI{50}{\percent}$ overlap were used, where the size of the smallest interrogation window was determined such that at least 10 particle images were present per interrogation area \cite{scharnowski2020}. The spatial resolution of the PIV setup is of order $\SI{0.02}{\milli\meter/pixel}$, resulting in a single velocity vector every $\SI{0.48}{\milli\meter}$, based on the smallest interrogation windows size. Following cross-correlation, semi-automatic data validation is performed by applying a local median filter to the vector field \cite{westerweel2005,thielicke2014}, followed by interpolation of the outliers.

\subsection{Measurement uncertainty}
\label{sec:exp_uncertainty}
\
The resolution of a schlieren system describes the minimum detectable ray deflection angle of the optical setup and represents a measure of the sensitivity of a given system. The contrast $C$ of schlieren images is defined by Settles \emph{et al.} as \cite{settlesSchlierenShadowgraphTechniques2001}:
\begin{align}
\label{eq:schlieren_contrast}
C =  \frac{\Delta E}{E} = \frac{2F \varepsilon_y}{a}.
\end{align}
Here, $\Delta E$ refers to the image intensity at a certain point, $E$ is the average background intensity, $\varepsilon_y$ is the angular ray deflection in the $y$-direction (cf. Eq.~\ref{eq:ray_deflection}), and $a$ is the unobstructed height of the source image on the cut-off plane. It is noted that the original equation quoted by Settles \emph{et al.}~\cite{settlesSchlierenShadowgraphTechniques2001} was multiplied by a factor of 2 to account for the location of the mirror at the radius of curvature in the current setup (see figure~\ref{fig:schlieren_system}).

Assuming that $\SI{50}{\%}$ of the source image were cropped by the knife edge at the cut-off, Eq.~\ref{eq:schlieren_contrast} yields for the minimal detectable ray deflection angle \cite{settlesSchlierenShadowgraphTechniques2001}:
\begin{align}
\varepsilon_{y, \mathrm{min}} =  
C \frac{a}{2F}
%= \SI{6.25e-6}{rad} \, \widehat{=} \, \SI{1.3}{arcsec}
\end{align}
For the present study, a circular pinhole of $\SI{300}{\micro\meter}$ in diameter was used, which at $\SI{50}{\%}$ cut-off results in an unobstructed image height of $a = \SI{150}{\micro m}$. On the assumption that, the minimal detectable intensity difference is $\SI{10}{\%}$ \cite{settlesSchlierenShadowgraphTechniques2001}, the minimal detectable ray deflection angle for the present schlieren setup is estimated as $\varepsilon_{y, \mathrm{min}} = \SI{6.25e-6}{rad} \, \widehat{=} \, \SI{1.3}{arcsec}$. As illustrated by means of Table~\ref{tab:deflection_angles}, this value for $\varepsilon_{y, \mathrm{min}}$ compares favourably to other schlieren setups found in the literature.

\begin{table}[t]
\centering
\caption{Examples of minimal ray deflection angles $\varepsilon_{\mathrm{min}}$ of other schlieren systems compared with this study.}
    \label{tab:deflection_angles}
    \begin{tabular}{@{}cc}
    \toprule
    Source & $\varepsilon_{\mathrm{min}}$ / arcsec \\ \midrule
    Kouchi \emph{et al.} \cite{kouchiFocusingschlierenVisualizationDualmode2015} &  17   \\
    Weinstein \cite{weinsteinLargefieldHighbrightnessFocusing1993} & 4, 16, 8, 24 \\
    Elsinga \emph{et al.} \cite{elsingaAssessmentApplicationQuantitative2004} & 4.13, 1.44, 2.68  \\
    Passmann \emph{et al.} \cite{passmann2020} & 25  \\
    Present study & 1.3  \\
   \bottomrule
    \end{tabular}

\end{table}

The computed value for $\varepsilon_{\mathrm{min}}$ represents the smallest theoretically possible value, i.e. the highest possible sensitivity, that can be achieved with a given schlieren system. In reality, geometric and optic imperfections of the individual system components, which are inherent to any actual optical setup, will increase the minimal detectable ray deflection angle \cite{settlesSchlierenShadowgraphTechniques2001}.
  
Estimation of PIV measurement error, in general, is a complex and non-trivial process \cite{raffel2018}. The measurement error is influenced by many parameters such as particle image size, noise level, and in- and out-of-plane loss of particles, to name only a few. In the present study the uncertainty in the computed particle displacement was estimated based on the data presented in \cite{thielicke2021}. Assuming a noise level of $0.015$ and a particle loss between the two images of $\SI{15}{\%}$ results in an average uncertainty in computed pixel displacement of  $\SI{\pm 0.2}{px}$, which translates to an uncertainty in flow velocity of approximately  $\SI{\pm 0.005}{\meter \second^{-1}}$.

\subsection{Simulation setup}

\begin{figure}[htbp]
    \centering
    \includegraphics[width=0.75\linewidth]{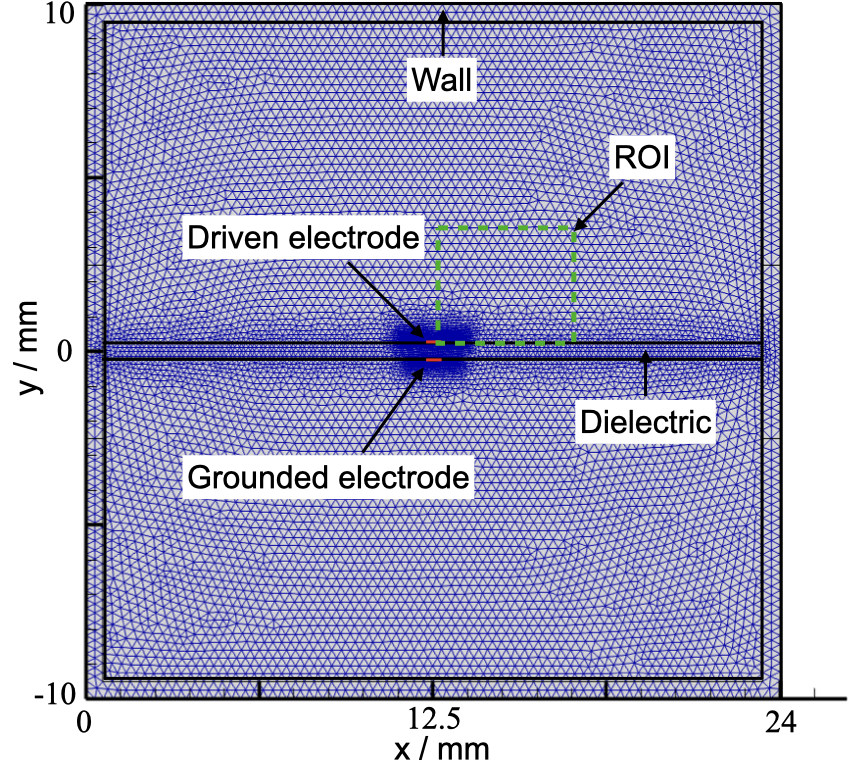}
	\caption{Schematic of the \textit{nonPDPSIM} simulation setup related to the single line electrode 1). The blue lines represent the numerical mesh.}
	\label{fig:sim_setup}
\end{figure}

In order to support the hypotheses formulated based on the sophisticated diagnostics methods described above, and to gain furthermore a deeper physical insight into the complex physics of the interaction between the dynamics of SDBDs and the initiated gas flow, numerical simulations are performed. We adopt Mark Kushner's simulation platform \textit{nonPDPSIM} which solves for the plasma and neutral transport in a fluid picture on an unstructured mesh allowing for a complex gas phase chemistry \cite{kushner2004modeling,kushner2005modelling}. It has been shown that \textit{nonPDPSIM} is well suited for simulating particularly dielectric barrier discharges in various scenarios\cite{tian2014atmospheric,kruszelnicki2016propagation,kruszelnicki2020formation,babaeva2010intracellular}. More specifically, in \textit{nonPDPSIM} Poisson's equation and transport equations are solved for the electric potential and for densities and momenta of charged and neutral particles, respectively. Additionally, the energy equation for electrons is consistently solved. Transport and reaction rate coefficients are obtained from a local solution of Boltzmann's equation. Furthermore, radiation transport is included. Details of \textit{nonPDPSIM} including equations and numerical methods are described in \cite{norberg2015formation}.

Our simulation scenario, for which we assume synthetic air N$_2$/O$_2$ = 80/20 at atmospheric pressure as the gas mixture, is shown in figure \ref{fig:sim_setup}. The unstructured mesh is clearly visible. In the centre of the simulation area, the mesh is much finer since this is the region of interest (the green dashed rectangle in figure \ref{fig:sim_setup}) where the plasma is ignited. The horizontal area in the centre represents the dielectric material with a thickness of 0.635 mm (aluminium oxide with a dielectric constant of $\varepsilon_r \approx 9$).  Due to numerical limitations, only a small region of the experimental SDBD setup is considered (24 x 20 mm$^2$). This region represents only one grid line at the top and bottom. Consequently, this simulation setup is used to compare the experimental results of the single line electrode 1). 

In our scenario, the bottom electrode is grounded. The red areas indicate their position. A voltage pulse of 12 kV with a slew rate of 1.2 kV/ns is applied to the top electrode. The pulse is on for 10 ns. During this period of time, the plasma dynamics are calculated. At the time of 10 ns, the plasma dynamics, i.e., all plasma related quantities as well as the transport and rate coefficients, are numerically frozen. These quantities are subsequently used as input parameters for the model of the neutral gas dynamics. Here we use compressible Navier Stokes equations using the SIMPLE algorithm for laminar gas flow. Whereas the plasma dynamics is solved on the nanosecond timescale, the neutral gas dynamics is solved on the millisecond timescale. This "Freezing" concept is introduced as a generic method to couple the fast timescale plasma dynamics and long timescale gas dynamics in one and the same simulation approach \cite{babaeva2018production,babaeva2023co2}.

\section{Results and discussion}
\label{chap:results}
\subsection{Experimental results}
\subsubsection{Electrical characterization}

\begin{figure}[hbt]
    \centering
    \includegraphics[width=0.65\textwidth]{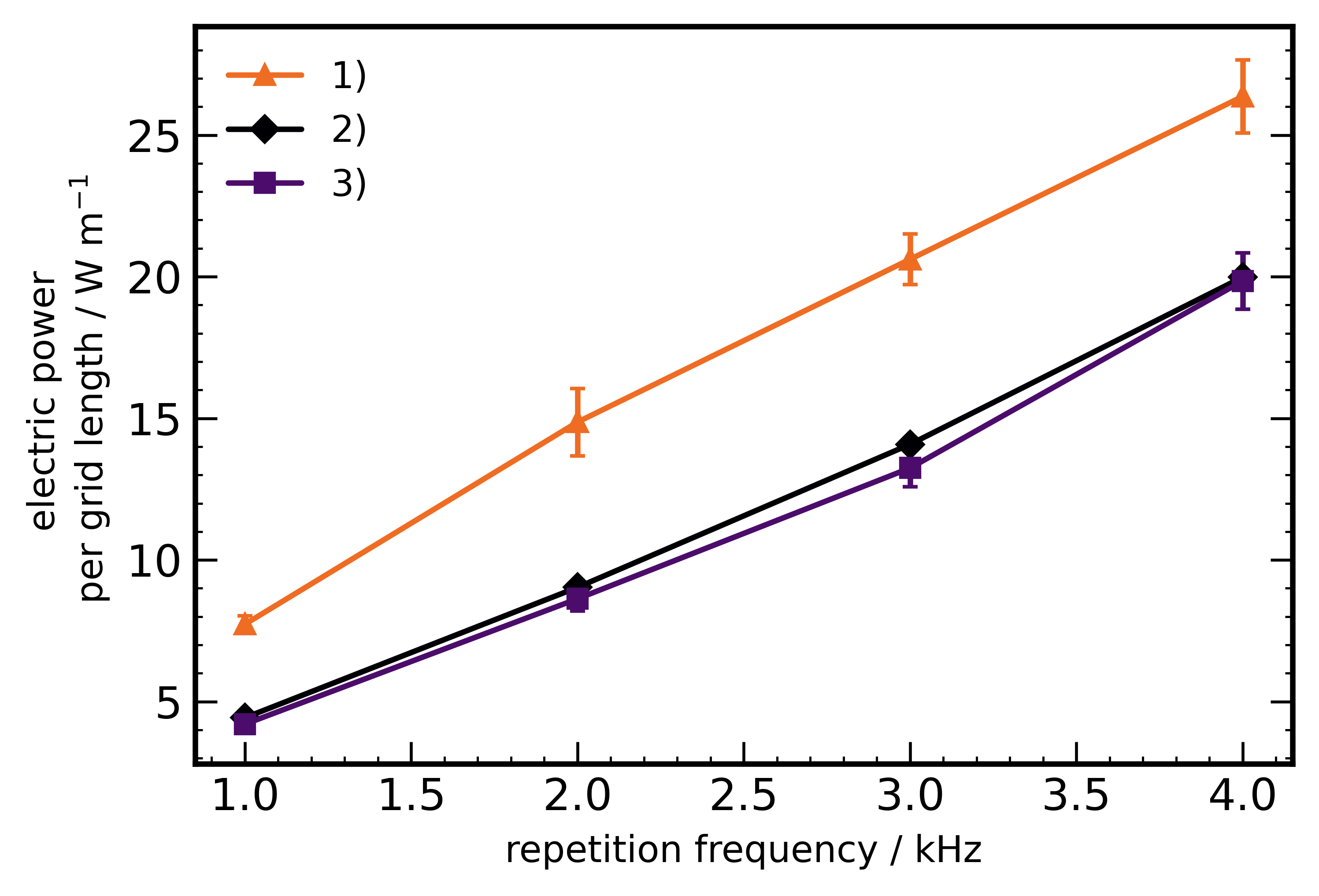}
	\caption{Dissipated electric power as a function of the repetition frequency for the three used electrode configurations 1)-3). The voltage was kept constant at \SI{11}{kV_{pp}}. The data for configuration 3) is extracted from Schücke \emph{et al.}, and it includes additional gas flow, which is not expected to change the U-I characteristics \cite{schuckeConversionVolatileOrganic2020}.}
	\label{fig:power}
\end{figure}

Figure \ref{fig:power} shows the dissipated electric power per grid length as a function of the repetition frequency of the discharge for all three used electrode configurations used in this study. The error bars represent the standard deviation obtained from reproduced measurements\cite{schuckeConversionVolatileOrganic2020}. The dissipated electric power rises linearly with the repetition frequency for all electrode configurations. Because the number of ignited discharge filaments depends on how often the applied voltage exceeds the breakdown voltage, a repetition frequency enhancement should lead to an increase of the filament number per time. As a result, both the current density and the dissipated power show a linear dependence on the repetition frequency, as confirmed by the experimental measurements. \\
Furthermore, it should be noted that variations in electrical capacitance arise due to changes in the electrode grid geometry among the different electrode designs. This disparity in capacitance has a significant impact on the system, particularly on the resonance frequency and the number of ignitions occurring within each pulse period. This variation in ignition events per pulse period is expected to influence the time required for the system to reach a stationary flow field. Exemplary waveforms recorded for the electrode configuration 3) (squared grid) can be found in Schücke \emph{et al.} \cite{schuckeConversionVolatileOrganic2020} and Böddecker \emph{et al.} \cite{boddeckerScalableTwinSurface2022b}. These waveforms are subjected to a fast Fourier transform, and the resulting resonance frequencies are summarized in Table \ref{tab:resonance}:

\begin{table}[ht]
\centering
    \caption{Resonance frequencies $f_{\mathrm{r}}$ for all used electrode configurations calculated with a fast Fourier transform analysis of the measured waveforms.}
    \begin{tabular}{@{}cc}
    \toprule
    electrode configuration & resonance frequency $f_{\mathrm{r}}$ / kHz \\ \midrule

    1) & 185  \\
    2) & 81 \\
    3) &  86 \cite{schuckeConversionVolatileOrganic2020}   \\

   \bottomrule
    \end{tabular}
    \label{tab:resonance}
\end{table}

While electrode configurations 2) and 3) exhibit a comparable resonance frequency due to their similar total gird area and geometry are similar, there is a significant difference of configuration 2) and 3 ) to the single grid line electrode 1). This result is expected, because the total grid line area is decreased in this case drastically, which leads to a reduction of the total capacitance. While a higher resonance frequency can lead to an increased number of ignitions per pulse period, the dissipated power per grid length is reasonably higher for the single grid line electrode while the voltage and the repetition frequency are kept constant.

\subsubsection{Transient start phase}

\begin{figure}[hbt]
    \centering
    \includegraphics[width=0.7\textwidth]{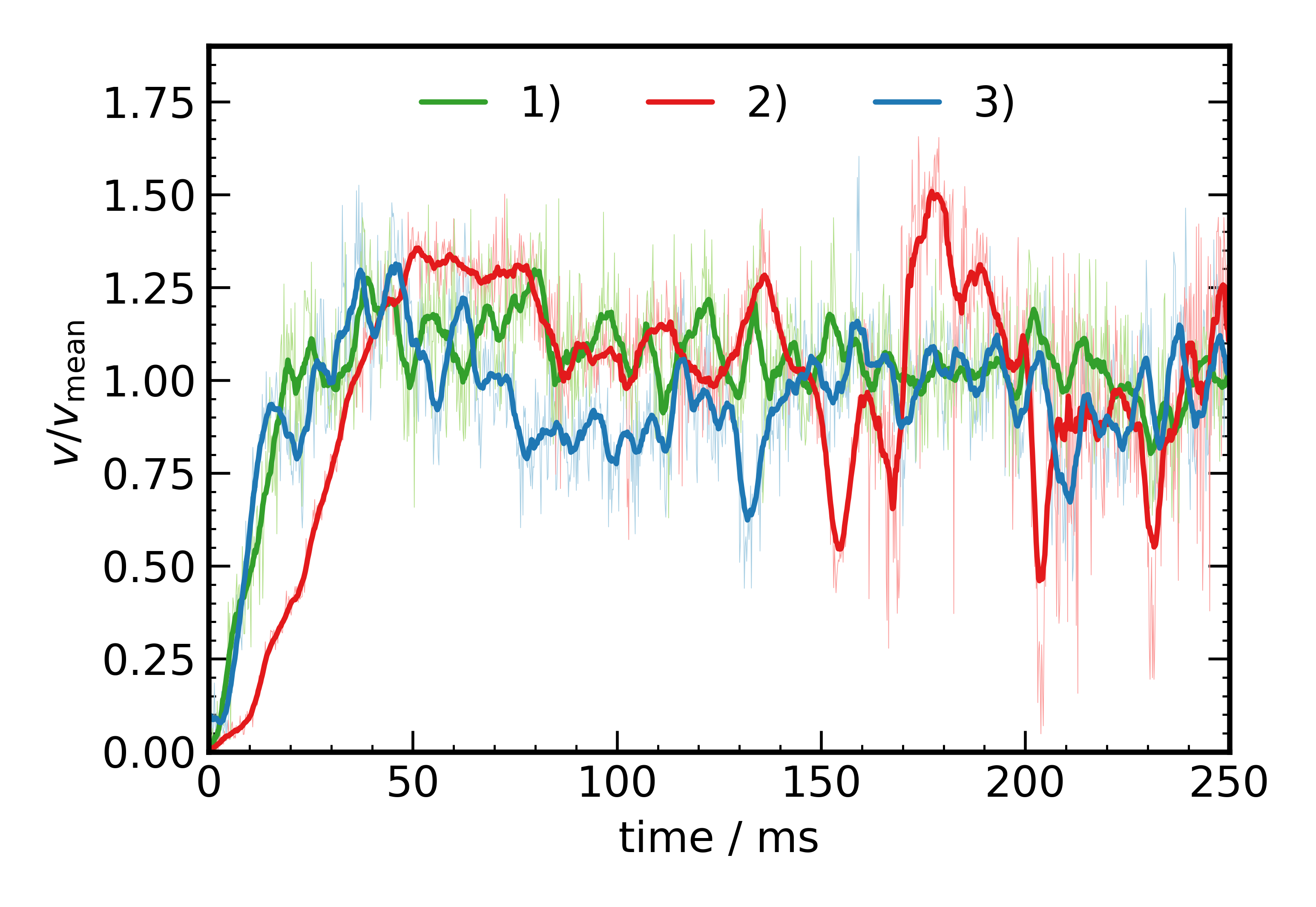}
	\caption{Normalized temporal velocity profiles for the single grid line electrode 1), the parallel grid line electrode 2), and the squared  grid line electrode 3) to the mean velocity, that is reached in the steady state case. Plotted are the calculated velocities from the PIV data and their corresponding moving averages, which take 20 samples into account, to enhance the visibility. The vertical velocity component $v$ is selected for all electrode configurations at a height of \SI{3.5}{mm} over the electrode grid, where only a vertical velocity component is found. The SDBDs are operated with a fixed voltage of \SI{11}{kV_{\mathrm{pp}}} and with a repetition frequency of \SI{4}{kHz}.} 
	\label{fig:start_phase}
\end{figure}

Figure \ref{fig:start_phase} shows the temporal development of the velocity in the three considered electrode configurations 1) to 3) within the first \SI{0.25}{s} following the initial ignition of the SDBD. The vertical velocity component $v$ at a single location, centrally above a grid line at a height of \SI{3.5}{mm}, is plotted versus time. For each electrode configuration the velocity $v$ is normalized by the corresponding time-averaged velocity $v_{mean}$ at that specific location. 
For all electrode configurations, $v/v_{mean}$ increases rapidly during the start phase, reaching a constant value in a subsequent steady state phase. However, the velocities display a relatively high variance. Figure \ref{fig:start_phase} enables the identification of a specific time point, denoted as $t_{\mathrm{start}}$, which marks the end of the transient start phase and the transition into the steady-state phase. Here, we have chosen to define $t_{\mathrm{start}}$ as the point at which $v$ reaches \SI{98}{\%} of $v_{\mathrm{mean}}$ for the first time. The calculated $t_{\mathrm{start}}$ values are summarized in table \ref{tab:t_start}.

\begin{table}[ht]
\caption{Duration of the transient start phase $t_{\mathrm{start}}$ for all three used electrode configurations 1) - 3). This point in time is defined by the first time of the velocity $v$ reaching \SI{98}{\%} of the corresponding $v_{\mathrm{mean}}$.}
\centering
    \begin{tabular}{@{}cc}
    \toprule
    electrode configuration & $t_{\mathrm{start}}$ / ms \\ \midrule
    1) & 40  \\
    2) & 60 \\
    3) &  40   \\
   \bottomrule
    \end{tabular}
    
    \label{tab:t_start}
\end{table}

Considering the resonance frequencies presented in table \ref{tab:resonance} for the different electrode configurations, one might expect variations in the time required to reach a steady state  for the single grid line electrode 1) relative to the other configurations 2) - 3), because their higher resonance frequency could lead to more discharge ignitions per pulse period. However, from figure \ref{fig:start_phase}, it can be observed that the values of $t_{\mathrm{start}}$ for each electrode configuration are very similar. Additionally, the start time for the single grid line electrode and the squared grid line electrode are identical. Only the start time for the parallel grid line electrode is \SI{50}{\%} larger relative to the others. From these observations, it can be concluded that the start time is not significantly influenced by the resonance frequency, or the configuration of the gridlines in general. In comparison, Whalley and Choi found a $t_{\mathrm{start}}$ of approximately \SI{100}{ms}, while their maximum velocity and applied repetition frequency is much higher \cite{whalleyStartingTravelingColliding2010}.
% Do we have to show here different start times for different frequencies also?

\subsubsection{Baseline design - single grid line electrode configuration}
\
%\begin{itemize}
 %   \item qualitative description (schlieren/PIV)
 %   \item validation case of experimental methods - comparision with DBD literature data
 %   \item charaterization (transient startup and steady state)
 %   \item velocity profiles
 %   \item FFT???
 %   \item avgerage velocity vs. frequency % %(schlieren (probably difficult...) and PIV!?)
\ %end{itemize}

%\begin{figure*}[hbt]
%    \centering
%    \includegraphics[width=0.95\textwidth]{graphs/SL2kHz_transientStartup_v1.pdf}
%	\caption{Single line transient start phase for a repetition rate $f=\SI{2}{\kilo\hertz}$ a)-f) and time-averaged steady state f)}
%	\label{fig:PIV_SL_trans_start}
% \end{figure*}

\begin{figure*}[hbt]
    \centering
    \includegraphics[width=\textwidth]{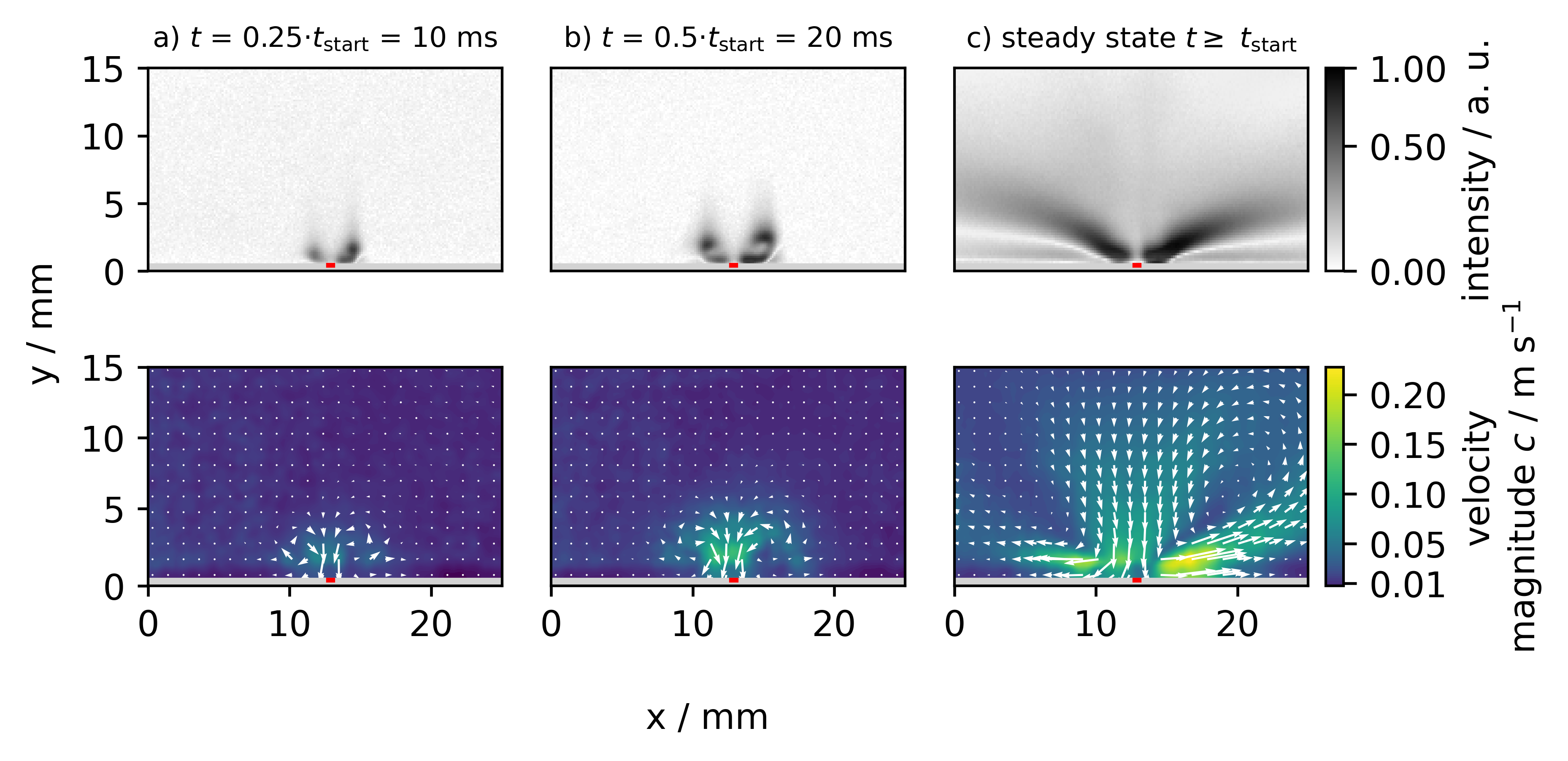}
	\caption{Transient start phase of the induced fluid flow of the single grid line electrode configuration 1) for a fixed repetition frequency of $f=\SI{4}{\kilo\hertz}$ at characteristic points in time a) -b) and for the time-averaged steady state c). The first image row shows schlieren images and the second row presents fluid flow velocity fields, evaluated from the PIV data. The gray horizontal bar and the red rectangles are artificially added and represent the electrode plate and the position of the grid line.}
	\label{fig:single}
\end{figure*}

% \begin{figure*}[hbt]
%     \centering
%     \includegraphics[width=0.95\textwidth]{graphs/02_SL2kHz_Macro_202208_BLprofiles_V2.pdf}
% 	\caption{Development of steady state velocity profiles in wall-normal direction for the single line electrode at a repetition rate of $f=\SI{2}{\kilo\hertz}$ \textcolor{red}{Achsenskalierung muss noch angepasst werden!}}
% 	\label{fig:SL_BLprofiles}
% \end{figure*}

Figure \ref{fig:single} shows schlieren images and velocity fields evaluated from the PIV data of the single line electrode at different points in time a) - b) after the initial ignition for a repetition frequency of \SI{4}{kHz}. Results in column c) show the time-averaged case for the time between $t_{\mathrm{start}}$ and the end of the recording ($\approx$ \SI{1.1}{s}) to visualize the steady state phase. The intensity of the schlieren images is normalized with respect to the maximum measured image intensity. To enhance the visibility of the schlieren structures, a single schlieren image captured before the initial ignition was chosen as the background image and subtracted from all subsequent schlieren images. The electrode's location is indicated by a horizontal gray bar in each image, while a red rectangle inside the electrode depiction shows the position of the grid line. \\

In figure \ref{fig:single} a) - c) clear schlieren structures are observed. In a), shortly after the initial ignition, small structures are detected, originating from the surface and extending towards the volume. Close to the surface, a small parallel structure is found that leads to a circular structure above it. The observed schlieren structures show a high degree of symmetry about the grid line (red rectangle), as was expected due to the symmetrical electrode configuration. Nevertheless, the right-hand side shows slightly higher intensities than the left-hand side. Close inspection of the electrode revealed a small in-plane misalignment between the upper and lower grid lines, due to manufacturing tolerances. A small asymmetry could lead to different electric field strengths on both sides of the grid line, explaining the observed intensity and size differences. \\

In figure \ref{fig:single} b), \SI{10}{ms} later, the schlieren structures retain their shape, but have increased in size. The vortices displayed here exhibit a  more closed structure, almost merging above the grid line. The small asymmetry is also visible here. In the steady state phase (see figure \ref{fig:single} c), approximately from \SI{50}{ms} on, the vortices cannot be identified clearly anymore. Two bright structures that originate at the grid line, and move out of the image, are dominant. Low intensity vertical structures are visible above the grid line. Because these structures are not visible in case a) and b) it can be stated that these structures represent the end of the vortices and do not originate from the grid line itself. The vortex increased in size above the dimensions of the imaging and cannot be resolved in total here. Parallel and close to the surface a second structure appears, which originates also from the grid line, but at a different angle to the dominant structure above it. Its total intensity is lower compared to the dominant structure above. The dark area between these structures suggests a  separation between these two structures. The previously discussed asymmetry leads to a smaller propagation angle on the right side and to a bigger area of high intensity. \\

The schlieren images can be compared to the fluid flow fields, which are positioned directly below them, to make statements about the qualitative behaviour and to validate the technique itself. While the  horizontal and vertical velocity components are visualized with the arrow field depiction, the velocity magnitude $c$ is plotted here as a filled contour plot. The PIV results were selected here for the same conditions in time as for the presented schlieren images to allow a direct comparison. \\
In the first two images for condition a) and b) the flow field far away from the electrode shows whether an external flow was present or not. Because no flow velocity could be evaluated in these regions, the enclosure is proven to protect the measurement region from external disturbances sufficiently. The first fluid flow field at condition a) is showing two small vortices originating from the grid line. The total resolved structure is bigger in size compared to the corresponding schlieren image. While the form of the outer structure of the vortices can be identified in the schlieren images, the velocity field directly above the electrode grid line is now also visible. There, the flow is directed from the gas volume towards the grid line. Close to the plate it turns to the sides and results in a closed loop for the two vortices equally. This behaviour could also have been expected for the found vortices in the schlieren images, because of the necessity of the flux continuity, but it is not visible here for condition a).\\
For condition b) the same description is valid as for condition a) and the structural size of the flow field is again bigger than in the schlieren image. While a slight asymmetry is visible in the flow field between the left-hand and the right-hand side of the grid line, the vortex centres are at similar mirrored spatial positions. An increase in the average velocity magnitude can be correlated with an intensity increase in the schlieren images. One main difference is, that the velocity magnitude $c$ and the schlieren intensity close to the surface do not match. That could lead to the fact, that not only fluid flow is causing the schlieren structures, but also temperature rises close to the grid line. \\
In the steady state phase c) it can be shown that the low intensity vertical schlieren structures above the grid line belong to the closing of the vortices. The flux originating from the volume towards the grid line reached its maximum size here. The asymmetry, that the intensity on the right-hand side of the grid line in the schlieren image is brighter and bigger in size, can also be found in the velocity magnitude $c$. Due to the smaller size of the vortex on the right-hand side, a fully closed structure is observable. Only close to the surface, high velocities in the range of \SI{0.2}{m \, s^{-1}} are achieved. Due to flux conservation the region of downward orientated velocity components is therefore bigger than the one which is flowing from the grid line to the sides. The schlieren structure parallel and close to the plate surface can not be found in the fluid flow. There are horizontal fluid flows visible but no structures with a discontinuity in vertical direction as dominant in the schlieren image. Therefore, this effect is probably caused by heating effects, that do not induce a relative high fluid flow. \\

The observations match with the results by Dickenson \emph{et al.} and Whalley and Choi with respect to the rotation orientation. The main difference is the temporal evolution for later points in time. While in Dickenson \emph{et al.} a jet like structure evolves due to the interaction of adjacent fluid flows, the results of Whalley and Choi measure also a detachment of the vortex from the surface, which is not visible here \cite{dickensonGenerationTransportReactive2018a, whalleyStartingVortexQuiescent2012}. 
%Because the size and form of the shown structures resembles the shown velocity field, it can be stated, that a fluid flow is mainly responsible for the refractive index gradient that causes the schlieren. This supports the assumption that the behaviour below and above the electrode is equal, because only anisotropic effects, as a temperature rise and its resulting buoyancy, are expected leading to differences. 

\subsubsection{Advanced design - parallel grid line electrode configuration}
\

% \begin{itemize}
%     \item qualitative description focusing on differences with respect to single line
%     \item charaterization (transient startup and steady state) ?
%     \item velocity profiles
%     \item avgerage velocity vs. frequency (schlieren and PIV!?)
% \end{itemize}

\begin{figure*}[hbt]
    \centering
    \includegraphics[width= \textwidth]{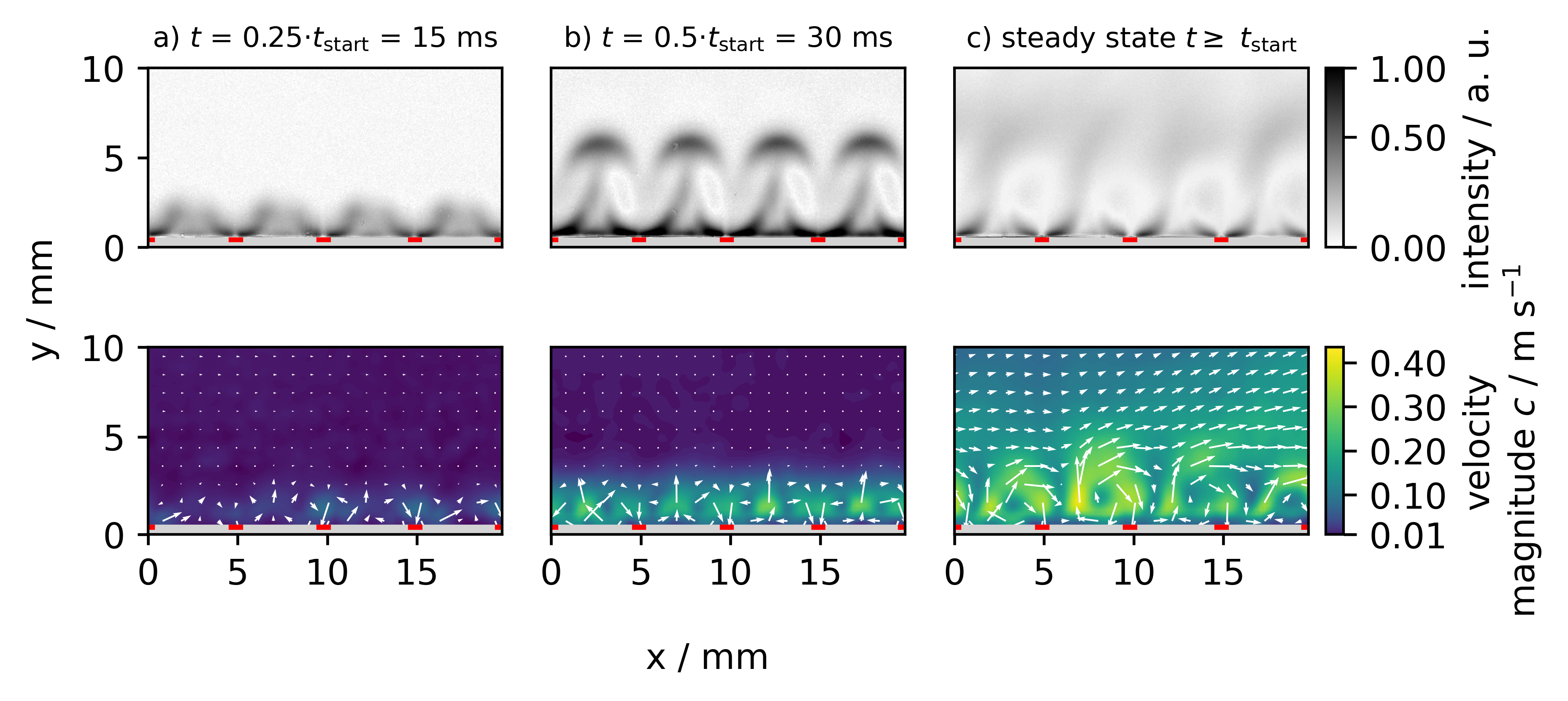}
	\caption{Transient start phase of the induced fluid flow of the parallel grid line electrode configuration 2) for a fixed repetition frequency of $f=\SI{4}{\kilo\hertz}$ at characteristic points in time a) - b) and for the time-averaged steady state c). The first image row shows schlieren images and the second row presents fluid flow velocity fields, evaluated from the PIV data. The gray horizontal bar and the red rectangles are artificially added and represent the electrode plate and the position of the grid line.}
	\label{fig:Schlieren_transient_parallel}
\end{figure*}

% \begin{figure*}[hbt]
%     \centering
%     \includegraphics[width=0.95\textwidth]{graphs/PL2kHz_transientStartup_v1.pdf}
% 	\caption{Parallel line transient start phase for a repetition rate $f=\SI{2}{\kilo\hertz}$ a)-f) and time-averaged steady state f).}
% 	\label{fig:PIV_SL_trans_start}
% \end{figure*}

Figure \ref{fig:Schlieren_transient_parallel} shows the temporal development of the recorded schlieren structures and the evaluated PIV data for the parallel grid lines electrode in the same way as in figure \ref{fig:single}. While in a) - b) definite schlieren structures are observable on the whole image, in c) only defined structures close to the electrode surface are remaining. Additionally, the intensity close to the surface decreases from condition b) to c). The images illustrate that the schlieren structure differs from the one with the single grid line electrode due to interactions between opposing flow field generated by discharges at adjacent grid lines. No closed vortices as for electrode configuration c) are visible here. In image b) the visibility of the schlieren structures is high, and all structures look equal. An angulated schlieren structure that originates from the grid line to the left and the right is observed, that is similar to the structure recorded for the single grid line electrode. The main difference is that the colliding schlieren structures resulting in a cap-like structure that moves normal to the electrode surface. The intense interaction between the induced flows of neighbouring grid lines change the direction of the fluid flow, so that the resulting flow is directed normal to the electrode surface. In image c) this cap-like structure moved further and a low intensity region remains followed by a dark region, where no schlieren structures can be observed. An asymmetry is observable in a), where the schlieren structures, that always originate from the left side of a grid line, look bigger and slightly higher than their opposing ones. Therefore, this asymmetry is similar to the one found for electrode configuration 1). While in b) always the left fluxes orientated upwards are slightly more dominant, the upper structures in the steady state phase seem to drift to the right side. This could be explained by the observed asymmetry in a) that could lead to an overall movement to the right as deviation of a perfect equilibrium of induced fluid flows. \\
In comparison to figure \ref{fig:single}, where the flow field structures have a bigger size than the schlieren structures, it is different for this electrode configuration. For condition a) the flow field is equally sized as the schlieren structures, and it shows an early stage of the vortex development as also the characteristic flow towards the grid lines from the gas volume, as observed for the single grid line electrode. At \SI{30}{ms} the schlieren structures are larger than the flow fields. Here, the assumption, that neighbouring opposing flows induce a flow normal to the electrode plate surface, can be confirmed. While the flow directly above the grid lines is directed towards the electrode surface, the flow between the grid lines moves upward after colliding, generating closed vortices. The fluid flow pattern is periodically repeated along the horizontal axis and the centre of the vortices have almost equal spatial positions, which results in a high symmetry. \\
The average velocity magnitude above the grid line at circa \SI{5}{mm} is slightly lower than the one at circa \SI{15}{mm}. While the flow field for the single grid line electrode at $t = \SI{20}{ms}$ has already surpassed a height of \SI{5}{mm}, the flow field is here more constricted, because even \SI{10}{ms} later, its height is lower than \SI{5}{mm}. In the steady state case, the velocities are substantially increased, and the vortex structures have grown in height. The asymmetry is more dominant in this case, and it is visible that also the far apart gas domain is influenced. There, a flux from the left to the right is induced. This observation matches with the drift, that is visible in the corresponding schlieren image. In contrast to the behaviour of the single grid line electrode, does this effect not increase the gas transport from the distant gas domain towards the discharge region. Here, only gas that is close to the vortices, at a height of roughly \SI{5}{mm}, is transported actively towards the grid lines. In comparison, the vortex size is similar to the one observed in Gilbart \emph{et al.} There, also small confined vortices are generated for a electrode width of \SI{1.6}{mm}, but no drift to one of the sides is visible. The rotation orientation is also clockwise, but the velocity magnitudes are faster despite the fact that our power density surpasses theirs significantly \cite{gilbartMutualInteractionMultiple2022}.

\subsubsection{Advanced design - squared grid line electrode configuration}

\begin{figure*}[hbtp]
    \centering
    \includegraphics[width = \textwidth]{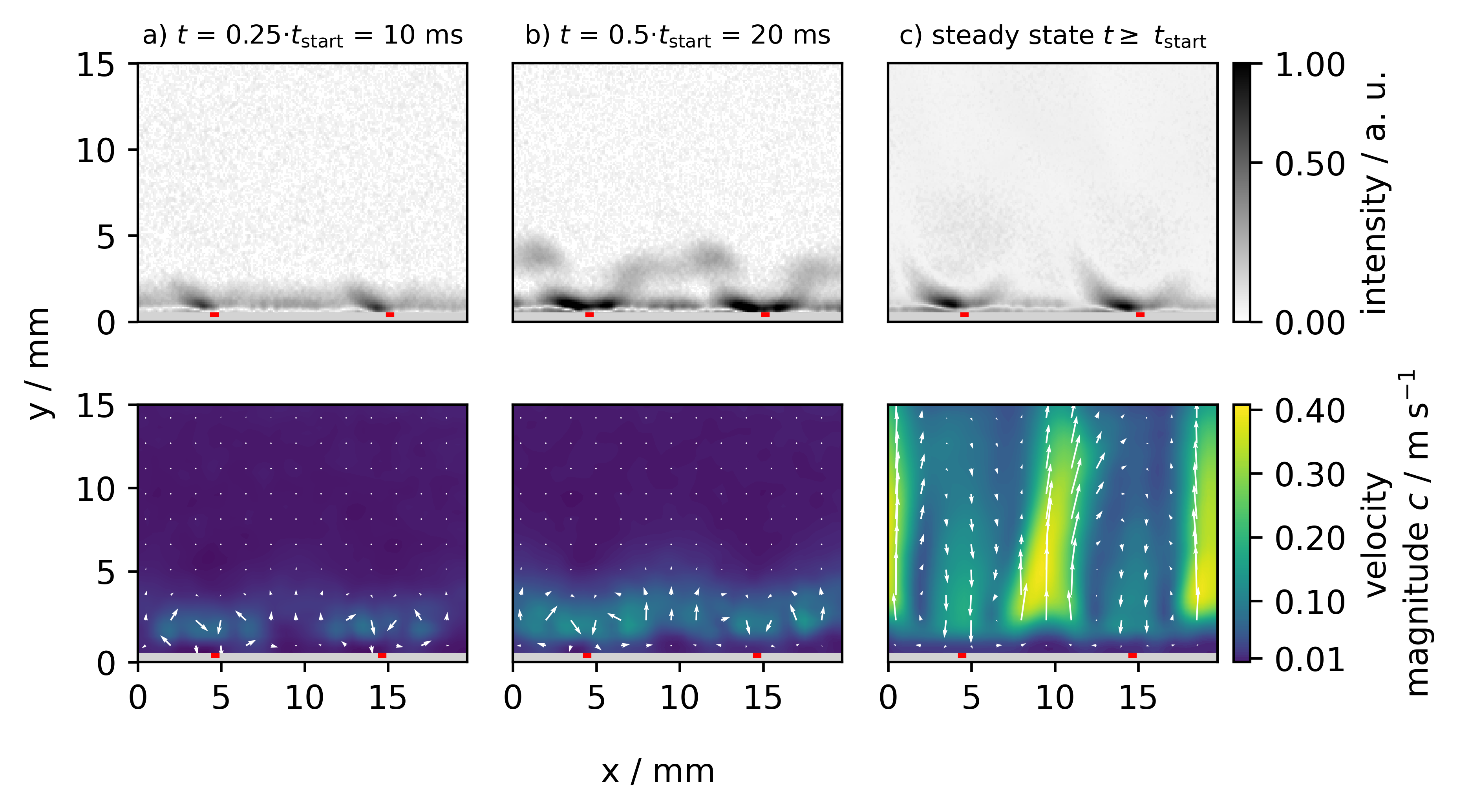}
	\caption{Transient start phase of the induced fluid flow of the quadratic grid line electrode configuration 3) for a fixed repetition frequency of $f=\SI{4}{\kilo\hertz}$ at characteristic points in time a) -b) and for the time-averaged steady state c). The first image row shows schlieren images and the second row presents fluid flow velocity fields, evaluated from the PIV data. The gray horizontal bar and the red rectangles are artificially added and represent the electrode plate and the position of the grid line.}
	\label{fig:Schlieren_transient_quadratic}
\end{figure*}

 Figure \ref{fig:Schlieren_transient_quadratic} shows the temporal development analogue to figures \ref{fig:single} and \ref{fig:Schlieren_transient_parallel} for electrode configuration 3). Because this electrode configuration has a similar geometry as electrode configuration 2) similarities are expected here. While the schlieren structures for condition a) and b) show a similar behaviour, that also matches the behaviour of the parallel grid line electrode, in c) no broad structure above the electrode is visible. The interacting colliding flows from neighboured grid lines are also visible here, but have a lower intensity and a more unsharp structure than for electrode configuration b). An asymmetry can also be found here. In contrast to the previous depicted electrodes, the structures have higher intensities and bigger sizes on the left-hand side of the grid lines, which is clearly visible in image a). In the steady state phase, where the maximum intensity is lower than in both other cases, the left-hand side is also brighter. The related PIV results show in a) and b) a similar behaviour as the parallel grid line electrode, but the general vortex structures are larger due to the doubled distance between the gridlines in the x direction. Here, no influence of the crossing grid lines are observable that would change this two-dimensional depiction of the flow field. For the steady state phase the general behaviour differs to the other designs completely. The middle structure at approximately \SI{10}{mm}, which is periodically repeated every \SI{10}{mm}, has the longest region of high velocity magnitudes compared to the other configurations. The vortices are substantially extended in vertical direction and a closing of the vortices is only visible in a narrow region at a height of roughly \SI{12}{mm}. Therefore, in contrast to the parallel grid line electrode, the vortex centre positions moved from condition b) to c) upwards. 
In comparison, Dickenson \emph{et al.} observed the same structural behaviour, where the adjacent fluid flows collide and form an upward directed fluid flow. They also showed that this results in enhanced transport of reactive species \cite{dickensonGenerationTransportReactive2018a}. We can therefore transfer this finding to our case.
The gas mixing is enhanced because the induced fluid flow is not constricted to a small region close to the electrode plate surface. Whether this effect originates from the doubled lattice constant or the crossing grid lines, which lead to a more three-dimensional behaviour, can not be answered here completely.

\subsection{Simulation results}
\label{chap:Simulation_Results}

\begin{figure*}[h]
    \centering
    \includegraphics[width=0.95\textwidth]{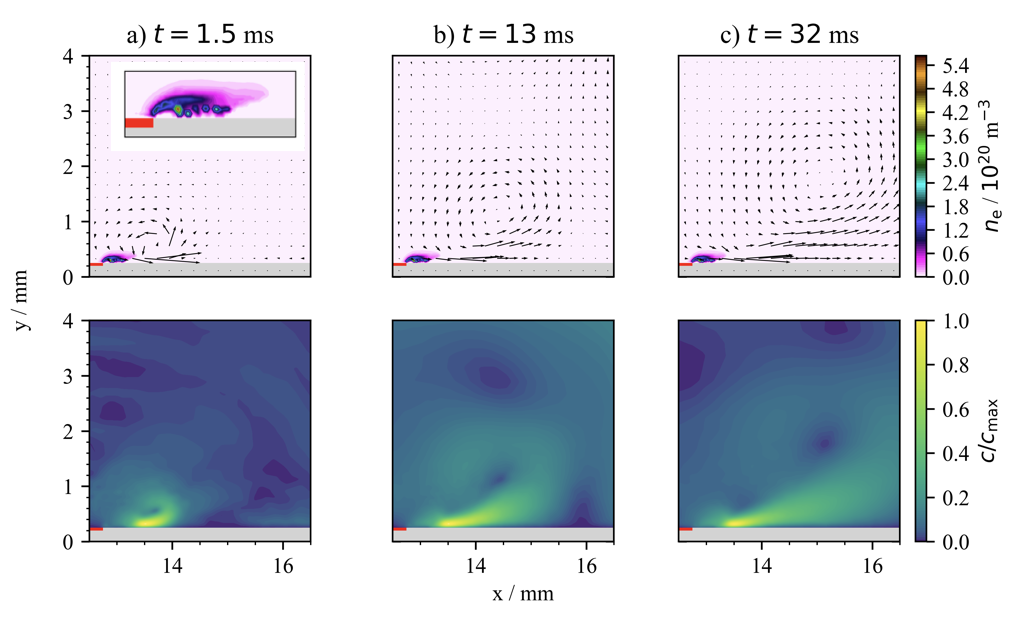}
	\caption{Simulation results for the region of interest at three different times. Top row: The electron density (colour plot) and the direction of the gas flow (black arrows). Bottom row: The normalized velocity of the gas flow.}
	\label{fig:Simulation_result}
\end{figure*}

In order to gain a better physical understanding of the interaction between the SDBD and the flow fields, 2D fluid simulations using \emph{nonPDPSIM} are performed as described above. As we are interested in the fundamental phenomena, we have concentrated only on the single grid line electrode (baseline design). It is worth mentioning that the high voltage pulse during the 10 ns leads to positive streamers at the top electrode -- as expected. The fully developed streamer on the right-hand side of the electrode can be seen in figure \ref{fig:Simulation_result} (top left) indicated by a high local electron density. Both, the electron density as well as the length of the streamer are in good agreement with Particle-In-Cell simulations done by Zhang et al. \cite{zhang2021computational}. The plasma quantities as well as the transport and reactions parameters are frozen at 10 ns and fed into the fluid model of the neutral gas dynamics. The neutral gas dynamics itself is solved on its own millisecond timescale. Figure \ref{fig:Simulation_result} shows the simulation results for three different instances of time on the fluid timescale (a) $t=1.5$ ms, b) $t=13$ ms, and c) $t=32$ ms). 

It is clearly visible that the streamer initiates a complex dynamics in the gas flow. Vortices are initiated, indicated by black arrows in \ref{fig:Simulation_result} (top row). This phenomenon can be understood having in mind that a positively charged streamer head attracts electrons which leads to an ionization wave in front of the streamer from the electrode to the final position of the streamer. At the same time the positive streamer head (formed by ions) transfers momentum to the neutral background gas. This interaction between the discharge and the neutral gas is consistently taken into account in the simulation model. As a result, a local electrohydrodynamic force (EHD force) at the streamer head transfers continuously energy to the neutral gas. \cite{dickensonGenerationTransportReactive2018a,boeuf2007electrohydrodynamic} This leads ultimately to the generation of vortices in the fluid flow. Due to the continuous energy transfer, the vortex formation increases with time. 

The bottom row of figure \ref{fig:Simulation_result} shows that the vortex forms with a counterclockwise orientation on the right-hand side of the top electrode. At the left-hand side -- which is not shown here -- it forms with a clockwise orientation. Therefore, the gas flows downwards towards the electrode. This is similar to the results from PIV and schlieren measurements shown in \ref{fig:single}.

Although the simulation results are in qualitative agreement with the experimental findings, the absolute values of the calculated gas velocities differ from the measured ones significantly. Here, the maximum velocity in the simulation close to the streamer reaches around 150 m/s, while the velocity in the vortex region is around 20 m/s. Since the region of maximum velocity is relatively small it is difficult to access directly by PIV in the experiment, therefore the velocity in the wider vortex region is a better point of comparison. Comparing experiment and simulation in this region gives a discrepancy in absolute velocity of around two orders of magnitude. This discrepancy can be explained by the fact that in the simulation the neutral gas "sees" constantly a fully developed streamer due to freezing the plasma dynamics, whereas in the experiment the neutral gas is affected be pulsed streamers. In the experiment the momentum transfer from the streamer to the neutral gas can be understood as an averaged momentum transfer. Therefore, an averaged local electrohydrodynamic force produced by the streamer is active, in contrast to the simulation. The next step of improving  the simulation is to implement multiple pulses and on- and off times for the plasma in order to allow for more realistic momentum transfer from the streamer to the neutral gas.

%Zhang et al. \cite{zhang2021computational} investigated the dynamics of the streamers at the via PIC/MCC 

%\begin{figure}[!hbt]
 %   \centering
%    \includegraphics[width=0.95\linewidth]{graphs/Fig_Sim_ElectronDensity.png}
%	\caption{The electron density in cm$^{-3}$ at a specific time.}
%	\label{fig:sim_edens}
%\end{figure}

%\begin{figure}[!hbt]
%    \centering
%    \includegraphics[width=0.95\linewidth]{graphs/Fig_Sim_ElectronDensityZoom.png}
%	\caption{The electron density in cm$^{-3}$ at a specific time. The white arrows represent the gas flow.}
%	\label{fig:sim_edens_zoom}
%\end{figure}

%\begin{figure}[!hbt]
%    \centering
%    \includegraphics[width=0.95\linewidth]{graphs/FigSimVelocityMag.png}
%	\caption{}
%	\label{fig:sim_velocity_zoom}
%\end{figure}

\section{Conclusion and future work}
\label{chap:conclusion}
In this study, induced fluid flows from surface dielectric barrier discharges with different electrode geometries were characterized with particle image velocimetry and schlieren imaging. A 2D fluid simulation with \emph{nonPDPSIM} allowed the connection between the streamer dynamics with the resulting flow field and schlieren structures. \\
For the most basic electrode configuration, the single grid line electrode, the development of the vortices is investigated. While vortices are visible in the schlieren images as well as in the flow field depiction, the flow field results show a higher influence on a larger space. The parallel grid line electrode results show confined vortices close to the electrode surface, that enhance the gas mixing locally. The interaction between neighbouring induced flows leads to velocities of roughly \SI{0.4}{m \,s^{-1}} which are twice as fast as in the single grid line configuration case. The squared grid line electrode does not show as dominant schlieren structures as in the parallel grid line electrode case, but the vortices are extended normally to the electrode surface. Here, compared to the other cases, the gas mixing is the strongest, because a larger gas volume is influenced. Probably, the  doubled grid lattice constant is the main influence here. Because this electrode configuration was used in the previous gas conversion studies \cite{schuckeConversionVolatileOrganic2020, petersCatalystenhancedPlasmaOxidation2021} and especially in the upscaled system \cite{boddeckerScalableTwinSurface2022b} the found conversion was surprisingly high, the enhanced gas mixing can be identified as one of the main effects.
\\
For further investigation, new measurements in flowing air will be conducted to complement the connection to the treatment of contaminated air for the application. %The influence of the gas temperature, electrode grid geometries, and opposing SDBDs in a multi-electrode system with variable wall- and electrode distances are also of interest. 
The results have shown, that the grid lattice constant is an important parameter for gas mixing. There the simulations enable the investigation of an optimum grid distance. The influence of different voltage waveforms for the gas treatment application can be also investigated resource saving with the simulation. 
%Using the 2D fluid simulation, the detailed interaction of the streamer with the background gas will be investigated. The focus will be on the electrohydrodynamic force (ion-wind) in order to investigate the momentum transfer and the resulting vortices. 
Additionally, 0-dimensional chemistry models, as used in Schücke et al. \cite{schuckeAnalysisReactionKinetics}, where diffusion and drift are used as main gas transport mechanisms, can be further enhanced.

%The results of the simulations could be used to set up a new hybrid simulation, which can also cover the temporal development of the fluid structures. Also machine learning approaches could be implemented here.

% \begin{flushleft}
% \section*{ORCID IDs}
% A Böddecker \url{https://orcid.org/0000-0002-2216-9515}  \\
% M Passmann \url{https://orcid.org/0000-0002-2403-6353} \\
% S Wilczek \url{https://orcid.org/0000-0003-0583-4613} \\
% R Skoda \url{https://orcid.org/0000-0003-3289-083X} \\
% I Korolov \url{https://orcid.org/0000-0003-2384-1243} \\
% T Mussenbrock \url{https://orcid.org/0000-0001-6445-4990} \\
% A R Gibson \url{https://orcid.org/0000-0002-1082-4359}
% \end{flushleft}

\section*{Acknowledgements}
This study was funded by the German Research Foundation (DFG) via projects A7 and A5 of the collaborative research centre SFB 1316 "Transient atmospheric pressure plasmas - from plasmas to liquids to solids". Additionally, it was supported by the Federal Ministry for Economic Affairs and Climate Action on the basis of a decision by the German Bundestag with the project “PLASKAT”. The authors acknowledge helpful discussions with Prof. Achim von Keudell. S.W. and T.M. gratefully thank Prof. Mark Kushner for providing the source code of \textit{nonPDPSIM} and his constant support in its use.
%\printinunitsof{in}\prntlen{\textwidth}

%\bibliography{sn-article.bbl}% common bib file
%% if required, the content of .bbl file can be included here once bbl is generated
%\input sn-article.bbl
%% BioMed_Central_Bib_Style_v1.01

%% BioMed_Central_Bib_Style_v1.01

\end{document}